\documentclass[manuscript,screen,nonacm]{acmart}
\acmConference[]{} % need to include this to suppress addresses footer

\newcommand{\workshopname}{GenAICHI: CHI 2024 Workshop on Generative AI and HCI}
\newcommand{\licensedetails}{Licensed under a Creative Commons Attribution 4.0 International License (CC BY 4.0). Copyright remains with the author(s).}
\newcommand\extrafootertext[1]{% this command adds a non-numbered footnote
    \bgroup
    \renewcommand\thefootnote{\fnsymbol{footnote}}%
    \renewcommand\thempfootnote{\fnsymbol{mpfootnote}}%
    \footnotetext[0]{#1}%
    \egroup
}

\AtBeginDocument{ % setup headers and footers
    \fancypagestyle{firstpagestyle}{
        \fancyhf{}
        \fancyfoot[L]{\sffamily\footnotesize \workshopname}%
        \fancyfoot[C]{\sffamily\footnotesize \thepage}
    }
    \fancyhf{}
    \fancyhead[L]{\sffamily\footnotesize\shorttitle}
    \fancyhead[R]{\sffamily\footnotesize\shortauthors}
    \fancyfoot[L]{\sffamily\footnotesize\workshopname}%
    \fancyfoot[C]{\sffamily\footnotesize\thepage}
    \extrafootertext{\licensedetails}
    }

\usepackage{xcolor}
\newcommand{\pheading}[1]{\vspace{2px}\noindent\textbf{#1}}

\begin{document}

%%
%% The "title" command has an optional parameter,
%% allowing the author to define a "short title" to be used in page headers.
\title{Can Nuanced Language Lead to More Actionable Insights? Exploring the Role of Generative AI in Analytical Narrative Structure}

%%
%% The "author" command and its associated commands are used to define
%% the authors and their affiliations.
%% Of note is the shared affiliation of the first two authors, and the
%% "authornote" and "authornotemark" commands
%% used to denote shared contribution to the research.

\author{Vidya Setlur}
\affiliation{%
 \institution{Tableau Research}
 \city{Palo Alto}
 \state{California}
 \country{USA}}

\author{Larry Birnbaum}
\affiliation{%
 \institution{Salesforce,}
 \city{San Francisco}
 \state{California}
  \institution{Northwestern University}
  \city{Evanston}
  \country{USA}
  }

\renewcommand{\shortauthors}{Setlur and Birnbaum, et al.}

\begin{abstract}
Relevant language describing trends in data can be useful for generating summaries to help with readers’ takeaways. However, the language employed in these often template-generated summaries tends to be simple, ranging from describing simple statistical information (e.g., extrema and trends) without additional context and richer language to provide actionable insights. Recent advances in Large Language Models (LLMs) have shown promising capabilities in capturing subtle nuances in language when describing information. This workshop paper specifically explores how LLMs can provide more actionable insights when describing trends by focusing on three dimensions of analytical narrative structure: \textit{semantic}, \textit{rhetorical}, and \textit{pragmatic}. Building on prior research that examines visual and linguistic signatures for univariate line charts, we examine how LLMs can further leverage the \textit{semantic dimension} of analytical narratives using quantified semantics to describe shapes in trends as people intuitively view them. These semantic descriptions help convey insights in a way that leads to a \textit{pragmatic} outcome, i.e., a call to action, persuasion, warning vs. alert, and situational awareness. Finally, we identify \textit{rhetorical} implications for how well these generated narratives align with the perceived shape of the data, thereby empowering users to make informed decisions and take meaningful actions based on these data insights.
\end{abstract}

%%
%% The code below is generated by the tool at http://dl.acm.org/ccs.cfm.
%% Please copy and paste the code instead of the example below.
%%


%%
%% Keywords. The author(s) should pick words that accurately describe
%% the work being presented. Separate the keywords with commas.

%% A "teaser" image appears between the author and affiliation
%% information and the body of the document, and typically spans the
%% page.

\begin{teaserfigure}
\setlength{\fboxrule}{0.002pt} % Adjust the thickness as needed
\setlength{\fboxsep}{1pt}    % Optional: sets the separation between the box and content
\color{lightgray}  
  \fbox{\includegraphics[width=.328\columnwidth]{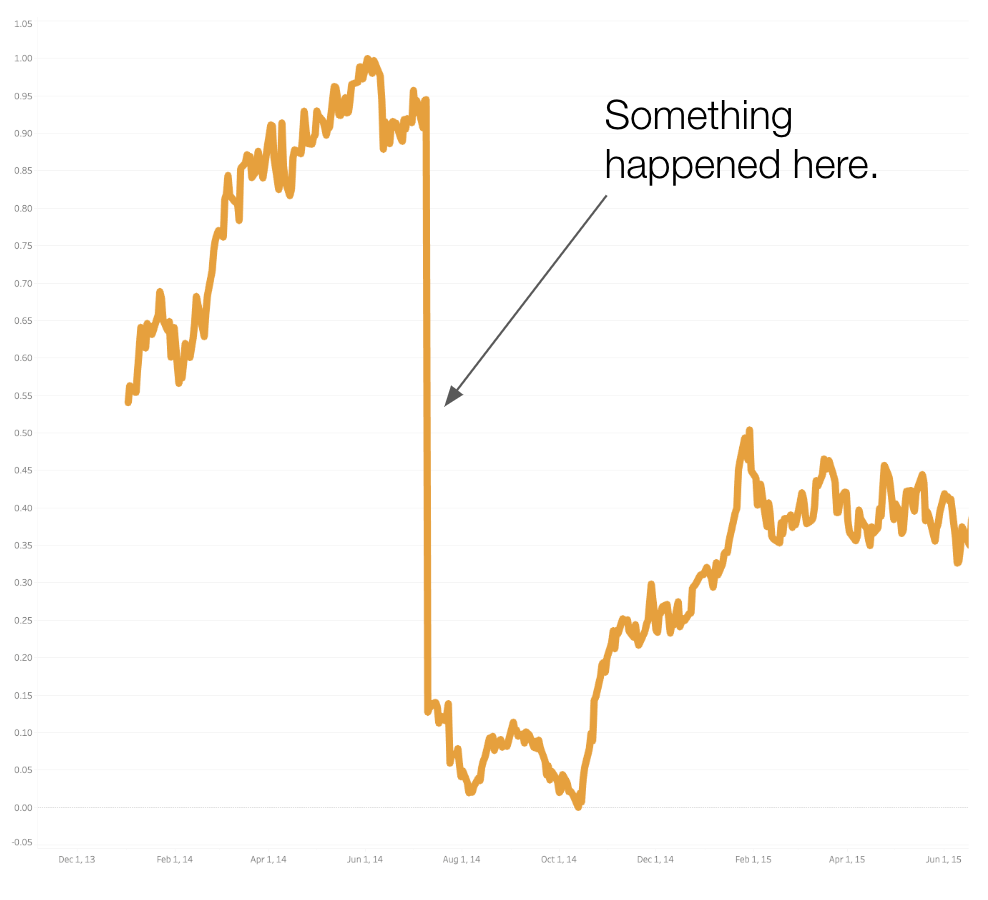}}
  \hspace{4mm}
   \fbox{\includegraphics[width=.45\columnwidth]{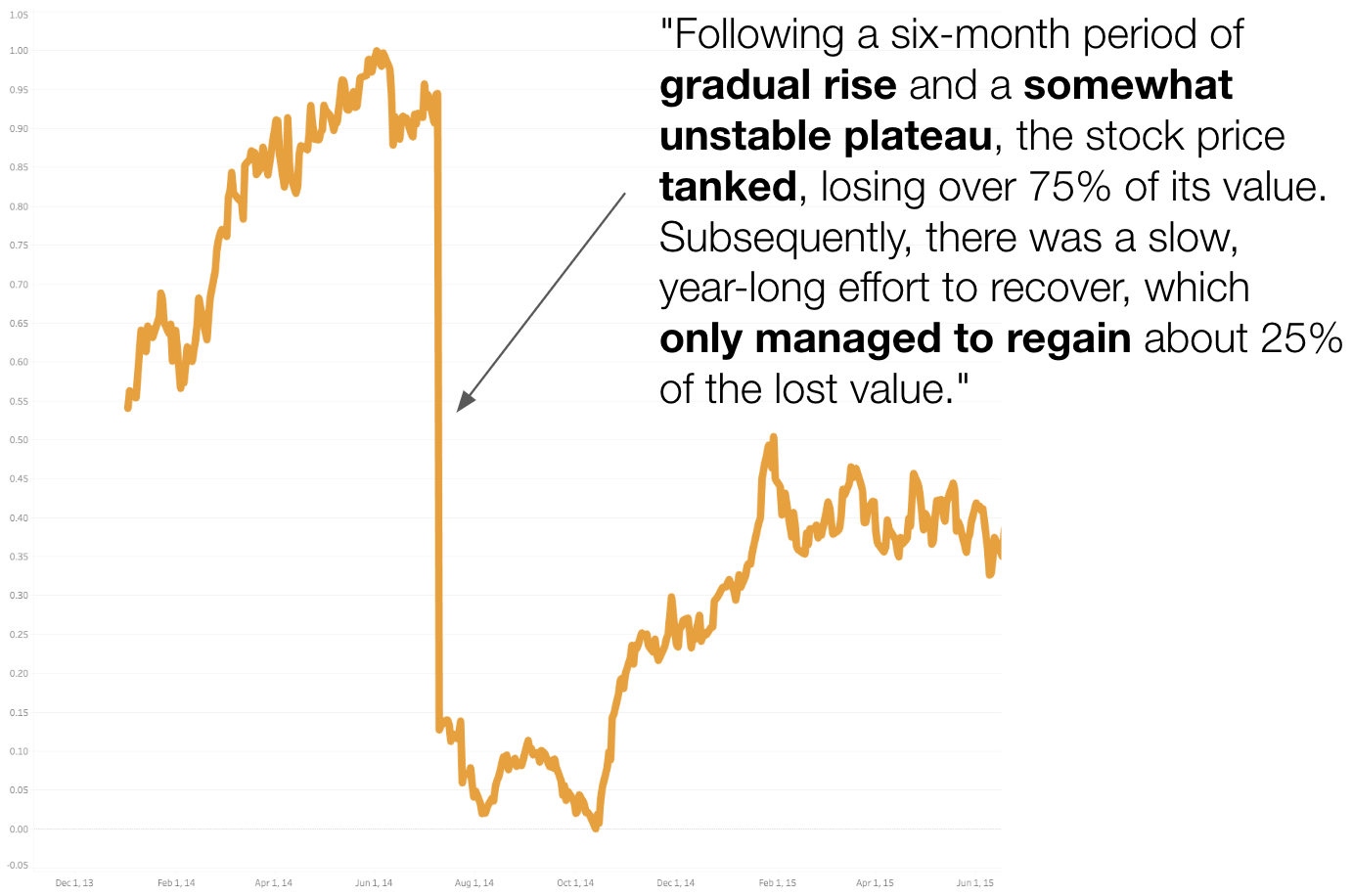}}
  \caption{Left: Relevant language describing visual features can be useful for authoring summaries to help with readers’ takeaways and provide context to what is occurring at the feature. Right:  A GPT-4-generated summary that incorporates nuanced language to draw a reader’s attention to specific portions of the text that are important to the overall takeaway of the intended message.
}
  \Description{Enjoying the baseball game from the third-base
  seats. Ichiro Suzuki preparing to bat.}
  \label{fig:teaser}
\end{teaserfigure}

% \received{20 February 2007}
% \received[revised]{12 March 2009}
% \received[accepted]{5 June 2009}

%%
%% This command processes the author and affiliation and title
%% information and builds the first part of the formatted document.
\maketitle

\section{Introduction}
Relevant language describing visual features in charts can be useful for authoring summaries about the data to help with readers’ takeaways and provide some context to what is happening in a data trend. However, the language employed in the recommended captions and annotations tends to be simple, ranging from specifying the domain, axes, and encodings to providing statistical information (e.g., extrema) describing specific marks in the chart. 

The use of more nuanced language is crucial in data analysis to convey the subtleties and complexities inherent to patterns in the data. This precision in language facilitates a deeper understanding and leads to more effective informed decision-making~\cite{segel:2010,lee2015more}. Hedging is a common communicative strategy used in language to emphasize and bring a reader’s attention to specific portions of the text that are important to the overall takeaway of the intended message~\cite{lakoff1973hedges,hyland1998hedging}. The proliferation of Generative AI along with Large Language Models (LLMs) amplifies this capability by analyzing vast amounts of data and uncovering patterns, trends, and correlations that might be missed in manual analysis~\cite{frieder2023mathematical}. Furthermore, LLMs can simulate various scenarios and predict outcomes, enabling analysts to explore a wider range of possibilities and make more strategic decisions. In this paper, we explore how LLMS such as GPT-4~\cite{openai2023gpt} could be employed to further enrich analytical narratives to elicit effective takeaways and action by the intended reader. We specifically explore the enrichment of analytical narratives across semantic, rhetorical, and pragmatic dimensions to further improve the clarity and impact of data-driven storytelling in the analytical decision-making process.

\section{Related Work}
Natural language has been used to help facilitate both language interaction and generation to help people reason with data~\cite{datatone,thoughtspot,ibmwatson}. Systems like Eviza~\cite{eviza} and Analyza~\cite{analyza} incorporate contextual inferencing capabilities to enhance user interactions. Other systems, such as Evizeon~\cite{hoque2017applying} and Orko~\cite{orko}, explore pragmatics within analytical conversations by leveraging contextual understanding. Flowsense~\cite{flowsense} enables NL-based interactions in dataflow systems. However, existing NLIs primarily focus on analytical inquiry and do not adequately address the semantic, pragmatic, and rhetorical dimensions required for describing insights in trends, which our work aims to address using GPT models.

iGraph~\cite{ferres2006igraph} supports querying trends but offers limited support for interpreting their semantic features, particularly in quantifiable properties like slope features. Hoque et al. conducted a comprehensive survey of chart question-answering systems, identifying opportunities for more sophisticated models leveraging language and semantics to support data exploration~\cite{hoque2022cqa}. Olio, a hybrid search system, combines semantic Q\&A search with document-based exploratory search but lacks robust support for querying specific semantics related to trends~\cite{olio}. Our work extends beyond existing trend querying systems by leveraging GPT models to support the exploration of trends with a comprehensive understanding of their semantic, pragmatic, and rhetorical dimensions, thereby enhancing insights derived from visual data analysis.

\section{Analytical Narrative Structure for Describing Data Trends}
Analytical narrative structure encompasses the organization and presentation of findings to effectively communicate the story within the data~\cite{statsmonkey}. The narrative should carefully incorporate descriptive language that accurately reflects the nuances of the observed trends, ensuring precision and clarity in conveying the message. We now discuss research opportunities that LLMs afford for the integration of semantic precision, pragmatic relevance, and rhetorical persuasion in this analytical narrative structure. We believe such a framework could provide a basis for effectively describing insights in trend data, driving understanding, decision-making, and action.

\subsection{Semantic Dimension}
Specific word choices play a crucial role in how the meaning or \textit{semantics} of trend shapes and patterns are perceived and understood. The language used to describe data trends significantly influences the audience's perception and interpretation, making the precision and subtlety of language an essential aspect.  Accurate descriptions of trend shapes, such as ``spike,'' ``drop,'' ``plateau'' or ``fluctuation,'' are not merely descriptive but carry connotations that impact the audience's understanding. These terms can alter the perceived significance or implications of a trend, necessitating the use of precise language that aligns with the data's narrative. This premise aligns with the recent work of Bromley and Setlur~\cite{bromley2023difference}, who explored the importance of labeling semantic visual features in data analysis to facilitate a more nuanced understanding of trends. Their approach identifies subtle semantic differences in data trends and integrates the the collected dataset with LLMs for textual summary generation. In addition, contextualizing data within a \textit{specific} domain is also helpful in making insights more meaningful. For example, a flat trend in the financial sector may be called `constant,' while a more appropriate label could be `unchanged' for weather forecasting. Incorporating LLMs to generate these bespoke semantics could further augment the semantic dimension of analytical narrative structures for describing trends. These models generate nuanced descriptions, identify contextually relevant linguistic patterns, and suggest alternative phrasings based on the domain that better capture the subtleties of the data.

\subsection{Rhetorical Dimension}
The rhetorical dimension focuses on conveying the semantics of these trend descriptors to the reader such that the analytical narrative can convey information that leads to either a call to action, persuasion, alert, or providing situational awareness. Language that provides effective rhetoric typically involves the nuanced usage of hedge words and connectives as the atomic units of communication~\cite{rhetoric:2014}. For instance, the choice between terms like ``fell sharply'' versus ``crashed'' is not just about word choice; it conveys different degrees of severity. Similarly, ``stagnant'' vs. ``stable'' also convey different expectations, with the former generally having a negative connotation, suggesting a lack of growth or change. In contrast, ``stable'' has a more neutral or positive connotation, indicating consistency and reliability.

LLMs could be appropriately prompted to emphasize certain aspects of data trends that align with a specific call to action. This process of appropriate narrative construction involves choosing words and constructing sentences that convey the appropriate level of urgency and importance. For data trends specific to a particular domain (e.g., financial reporting, public health data), GPT can employ language that includes explanations, contextual information, and possibly comparisons with past data or predictions for the future. Gricean principles, also known as cooperative principles, are a set of guidelines in the field of pragmatics for effective communication~\cite{Grice1975-GRILAC-6}. These principles include four Maxims of Quantity, Quality, Relation, and Manner to support these cooperative properties. The principles have been shown to be appropriate scaffolds within a GPT prompt to help generate an effective rhetoric. For example, in the generated insight, ``Revenue fell; however, drilling down further, the revenue from apparel decreased significantly,'' supports the Maxims of Manner (i.e., the narrative should be informative as is required) and Quality (i.e., the narrative should provide truthful and adequate evidence) by first stating the overall situation (i.e., revenue fell) and then adding a specific detail (i.e., the apparel revenue's significant decrease), which is directly relevant to the overall context (Maxim of Relation). The use of connectives, such as, `however' and phrasal transitions, such as `drilling down further,' avoids ambiguity and adds clarity to the narrative (Maxim of Manner).

\subsection{Analytical Connectives}
In addition to supporting cooperative language construct, we can draw upon several types of analytical connectives for generating rhetoric, especially in the context of data analysis and presentation~\cite{cardbook}. By explicating adding directives for the inclusion of analytical connectives in the prompt input to GPT, we can generate various forms of insight narratives that further enhance the rhetorical dimension, described as follows:

\pheading{Temporal connectives} involve connecting ideas or data points based on their temporal relationships. For example, temporal connectives are indicated in \textbf{bold} in this GPT-generated summary, ``The calm pattern in the stock market over the\textbf{ last month} with \textbf{minimal volatility}, reflects investor confidence and a lack of disruptive market news.''

\pheading{Part-whole relationships} focus on how individual data trends or elements contribute to or are a subset of a larger dataset. Phrases such as `a component of' or `an aspect of the broader trend' are useful in this context. For example, part-whole connectives are indicated in \textbf{bold} in this GPT-generated summary, ``The increase in overall profits this quarter was largely driven by a surge in online sales, which constituted \textbf{60\% of the total revenue}.''

\pheading{Comparison connectives} are used to draw parallels or highlight differences between different datasets or elements within a dataset by employing statements of comparison (e.g., `compared to,' `unlike,' `similar to') in the GPT-generated summary indicated in \textbf{bold}, ``While the company's revenue has increased, competitors have seen \textbf{similar growth}, implying that this trend is industry-wide rather than company-specific.''

\pheading{Roll-up or drill-down connectives} are used to describe trend data in terms of aggregation or disaggregation. A roll-up involves summarizing or aggregating data to a higher level (e.g., from daily sales to monthly sales), while a drill-down involves breaking data into more detailed components (e.g., breaking down annual revenue into product-wise revenue). Phrases like `summing up,' `in aggregate,' or `breaking down into' can be used to generate a summary such as, ``\textbf{Drilling down} into the total revenue, we observe that the North American market contributes 50\% of it, with a significant portion coming from online sales,'' indicated in \textbf{bold}.

\pheading{Normalization strategies} are used to describe how data can be adjusted or normalized to account for seasonal variations, cyclical trends, or different time scales. For instance, the summary with normalization phrases indicated in \textbf{bold} -  ``When \textbf{normalized year over year}, the company's revenue shows a consistent \textbf{seasonal pattern}, with spikes during the holiday season and dips in the second quarter.''

\subsection{Pragmatic Dimension}
The pragmatic dimension in describing analytical narratives for data trends focuses on the practical implications and actions that arise from the analysis. In other words, this dimension encompasses various aspects such as decision support, predictive analysis for future planning, risk management, effective resource allocation, policy and strategy formulation, performance improvement, and change management~\cite{pragmaticsbook}. We posit that LLMs can aid in supporting the pragmatic dimension by reassessing the analysis approach, evaluating existing methodologies, and suggesting improvements. For instance, in a scenario where a company observes a stagnation in market growth, the LLM can analyze existing market data and propose a shift in analysis towards emerging trends or unexplored customer segments. Additionally, when existing data is insufficient for informed decision-making, the LLM's role becomes crucial in identifying gaps and suggesting areas for additional data collection. For example, if a retail business notices a decline in sales, the LLM might suggest gathering data on recent market shifts, competitor strategies, or changing consumer preferences to gain a comprehensive view.

\section{Takeaways and Research Implications}
In summary, leveraging LLMs to provide semantic, rhetorical, and pragmatic dimensions for describing analytical narratives presents a promising avenue for future research. Specifically, the potential of GPT to generate hedge words and analytical connectives, supplied as a curated word bank in the prompt, opens up possibilities for creating more nuanced and contextually relevant narratives. These narratives, enriched with appropriate linguistic cues, can be effective for memorability and takeaways, thus eliciting a call to action. Investigating the impact of these GPT-generated narratives on decision-making could provide valuable insights into their practical applicability and effectiveness. 

\bibliographystyle{ACM-Reference-Format}
\bibliography{main}

\end{document}